\documentstyle[12pt]{article}
\input epsf.tex
\def\beq{\begin{equation}}
\def\eeq{\end{equation}}
\def\bea{\begin{eqnarray}}
\def\eea{\end{eqnarray}}
\def\bq{\begin{quote}}
\def\eq{\end{quote}}

\makeatletter
\def\vereq#1#2{\lower3pt\vbox{\baselineskip1.5pt \lineskip1.5pt
\ialign{$\m@th#1\hfill##\hfil$\crcr#2\crcr\sim\crcr}}}
\makeatother

\begin{document}

\begin{titlepage}
\begin{center}
December 1, 1998     \hfill    SLAC-PUB-8020\\
~{} \hfill hep-th/9812010\\

\vskip .6in

{\large \bf
Field Theoretic Branes and Tachyons of the QCD String}
\vskip 0.4in

Nima Arkani-Hamed and Martin Schmaltz

\vskip .2in
{\em SLAC, Stanford University, \\
Stanford, California 94309, USA}

\end{center}

\vskip.6in

\begin{abstract}
Dvali and Shifman have proposed a field-theoretic mechanism for
localizing gauge fields to ``branes" in higher dimensional spaces
using confinement in a bulk gauge theory.
The resulting objects have a number of qualitative features in
common with string theory D-branes; they support a gauge field and
flux strings can end on them. In this letter, we explore this
analogy further, by considering what happens when
$N$ of these ``branes" approach each other.
Unlike in the case of D-branes, we find a {\it reduction} of the gauge
symmetry as the ``branes" overlap. This can be
attributed to a tachyonic instability of the flux string stretching
between the branes.
\end{abstract}

\end{titlepage}

%\section*{Introduction}

Recently, Dvali and Shifman have
considered the possibility of trapping gauge fields on
$p$-branes with $p<3$ using confining
dynamics in a bulk 3+1-dimensional gauge theory \cite{Gia}.
These field theoretic branes are very interesting as
their higher dimensional generalizations can be used
to construct extensions of the standard model with extra
dimensions. In such models gravity and possibly some other
fields propagate in higher dimensional space-time whereas
the standard model matter and gauge forces are confined
to (3+1) dimensional branes \cite{ADD, extraD}. 
Apart from these potential phenomenological applications
field theoretic $p$-branes also provide a very interesting
background for studying the interplay of dynamics in various
dimensions. We are particularly intrigued by some obvious
similarities between these branes and D-branes \cite{joe}
in string theory:
apart from supporting a gauge field in their world-volume,
field theoretic branes also allow color flux-strings to end
on them \cite{schif}.
The aim of this letter is to further explore the analogy between
these branes in field theory and in string theory.  
More concretely, after reviewing the construction of Dvali
and Shifman and giving some simple generalizations, we consider
what happens when $N$ of these walls are brought on top of each
other. From the analogy
with D-branes one expects that modes of the QCD flux-string
become light and contribute to the effective $p$ dimensional
world-volume field theory. In the case of D-branes the lightest modes
are spin-1 fields (and their superpartners), and the $U(1)^N$ gauge
symmetry of $N$ widely separated D-branes gets enhanced
to $U(N)$.
In the case of the QCD string the masses of low-lying vibrational modes
decrease as
\beq
\label{mass}
m \propto L \ \Lambda^2
\eeq
when the length $L$ of a long flux tube is reduced. However which modes of
the QCD string become light for small $L$ turns out to be different.
We will argue from consistency of the $(2+1)$ dimensional
low energy field theory on the branes
that the lightest such mode is not a spin-1 field.
Instead we find a scalar whose mass
squared is positive for long stretched strings but becomes negative
for very small $L \sim \Lambda^{-1}$ where eq.\ref{mass} breaks down.
Thus for very small separation the scalar condenses and {\it reduces}
the gauge symmetry of the effective field theory on the branes.
This understanding of the reduction of gauge symmetry as branes
are brought in contact with each other from both a macroscopic effective
(2+1)-d theory as well as from a microscopic (3+1)-d picture with
QCD strings is our central result.
As a bonus we also find that field theoretic branes can be
connected by multi-pronged
flux tubes corresponding to ``baryonic'' QCD strings.

To get started we first review the argument of Dvali and Shifman
\cite{Gia} and give a simple generalization before moving on to consider
what happens when $N$ of these walls are brought on top of each other.

In the construction of our walls we will frequently assume that
a gauge group is broken in some region of space
but not in others. We will also have use for matter
fields which are very massive in the bulk but
light on the walls. In the discussion we will
assume that these effects have been arranged by coupling
the theory to a ``black box'' containing appropriate
very massive neutral and charged scalars with space
dependent vacuum expectation values\footnote{An example of
such a black box can be found in \cite{Gia}.}.

For simplicity, we first attempt to
localize a $U(1)$ gauge field to a region
${\cal W}$ in 3+1 dimensional space
between $0<z<l$, which on distances much
larger than $l$ would look like
a 2+1 dimensional wall supporting a $U(1)$
gauge field.

The most obvious idea is to arrange for the
$U(1)$ to be broken outside ${\cal W}$ giving
the photon a mass $M >>l^{-1}$,
but unbroken inside ${\cal W}$. Then, since
the photon is massive outside ${\cal W}$
but massless inside, one may think that there is a
massless electric photon in the $(2+1)$-d theory at long distances.
This is not the case. To understand this, note that the
region outside ${\cal W}$ is
superconducting while the region inside is normal vacuum.
Now place an electric test charge inside ${\cal W}$ and examine the
field strength at another point in ${\cal W}$ a distance $r >> l$
away; if there is a massless photon in the long-distance theory,
we should have a $(2+1)$-d Coulomb field in this regime.

% FIGURE 1
\begin{figure}[ht]
\centering
\epsfysize=1.9in\epsfbox{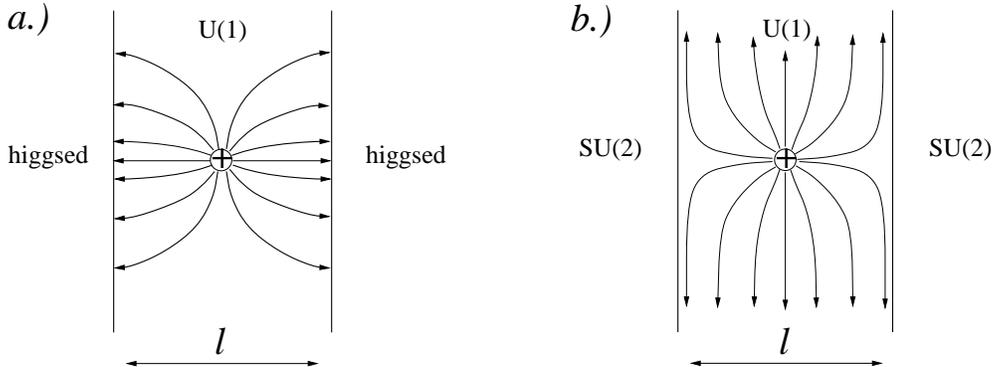}
\smallskip
\parbox{5in}{\caption{\it The figure labeled a.) depicts a domain
wall which has no massless electric photon trapped. This can be
seen from this figure by noting that the electric
field lines are screened by the superconducting Higgs vacuum in the
bulk. In figure b.) the bulk is in a confined phase and repels electric
flux. As a result, a massless photon coupling to electric charge is
trapped.
}}
\end{figure}

Since the
region outside is a conductor, however, the electric field lines
emanating from the test charge must end on and be perpendicular to the
boundary of ${\cal W}$, whereas in order to obtain a $2+1$
dimensional force law these field lines would have to be repelled
from the boundary. We can solve for the electric field using the
method of images, with an infinite number of image charges of
alternating signs. Clearly all the multipole moments vanish for such
a configuration, and we are left with an exponentially small field
for $r >>l$. Therefore, we conclude that there are no fields
lighter than the ultraviolet cutoff $l^{-1}$ of the (2+1)-d theory,
coupling to electric
charge. It is very easy to see (as we show in detail in the
appendix) that instead,
there is a tower of {\it massive} gauge fields with masses quantized in units
of $l^{-1}$.

This failure suggests the correct way to proceed, however. Suppose
that we instead place a magnetic charge $g$ inside ${\cal W}$.
Now, because of the Meissner effect in the superconducting region,
all the magnetic flux lines are repelled from the boundaries and
we recover the $(2+1)$-d magnetic Coulomb law. A trivial
application of Gauss' law yields the relationship between the
effective $(2+1)$-d magnetic charge $g_3$ and $g$:
\beq
\frac{1}{g^2_3} = \frac{1}{g^2} \times l
\eeq

Of course, we actually want to localize electric photons on the
wall, this can be accomplished by the `t Hooft-Mandelstam dual of
this superconducting picture. Suppose that we begin with a $(3+1)-d$
$SU(2)$ gauge theory, which is broken to a $U(1)$ inside ${\cal
W}$ by a very massive scalar in the adjoint representation of
$SU(2)$. The bulk theory is confining at the scale $\Lambda$ which we
take to be $>> l^{-1}$, whereas the the $U(1)$ inside ${\cal W}$
is free. If we now place an electric test charge inside ${\cal
W}$, confinement expels the electric field lines from the
bulk due to the dual Meissner effect, and we recover the $(2+1)$-d Coulomb
law for
the electric field. This successfully localizes a $U(1)$ gauge
field to a $(2+1)$-d wall in a $(3+1)$-d bulk.

There are obvious generalizations of this idea. Suppose we have an
$SU(N_c)$ gauge theory with $N_F >> N_C$ flavors, which are
given a very large mass outside ${\cal W}$ but are massless
inside. Then the outside theory is asymptotically free and confines.
The theory inside the region $\cal W$ is infrared free at
distances short compared to the wall thickness $l$ where the
coupling evolves according to the $(3+1)$-d renormalization
group equation. But at length scales long compared to $l$ the
theory on the wall is $(2+1)$-dimensional and the coupling evolves
according to the $(2+1)$-d renormalization group equation.
At the UV cutoff $l^{-1}$ of the low energy theory the 
(2+1)-d gauge coupling is matched to the higher dimensional
coupling as
\beq
g^2_3(\mu = l^{-1}) = \frac{g^2_4(\mu=l^{-1})}{l}
\eeq
By the same argument as for the $U(1)$ case above, this
localizes an $SU(N_c)$ gauge theory on the $(2+1)$ dimensional wall.
Notice that unlike the $U(1)$ case, this (2+1)-d theory also
confines; however the confinement scale is $\sim g^2_3$
which can be much smaller than the cutoff $l^{-1}$ if $g^2_4$
is small. This is easy to arrange since the 3+1-d theory inside ${\cal W}$
can have a small gauge coupling at its UV cutoff and gets
(logarithmically) weaker as it is scaled into the IR towards $\mu =
l^{-1}$. Therefore, there is a range of energies $g^2_4/l < E
<1/l$ where we can have an unconfined (2+1)-d $SU(N_c)$ gauge theory.
In this manner it is possible to engineer a large variety of field
theoretic branes with different gauge theories living on them.

What happens if we
move an electric charge from the wall into the confining bulk \cite{schif}?
The confinement tries to expel the electric field lines, but since a
net flux of electric field must be present at large distances by
Gauss' law, a string of electric flux forms between the charge in the bulk
and the wall as in figure 2.
Thus, these walls have a second qualitative feature
in common with D-branes: strings can end on them \footnote{Note
that our electric flux strings ending on a wall of unconfined gauge
field are the electric-magnetic dual to cosmic
strings (with their associated magnetic flux)
ending on a domain wall of unbroken gauge field
as described for example in \cite{sean}.
The microscopic physics allowing strings to end on domain walls here is
different from the physics allowing strings to end on domain walls
in $N=1$ supersymmetric QCD \cite{MQCD,Gia2}.
For a recent discussion of domain walls
in softly broken $N=2$ SUSY gauge theories, see \cite{Kapl}.}.

% FIGURE 2
\begin{figure}[ht]
\centering
\epsfysize=2.in\epsfbox{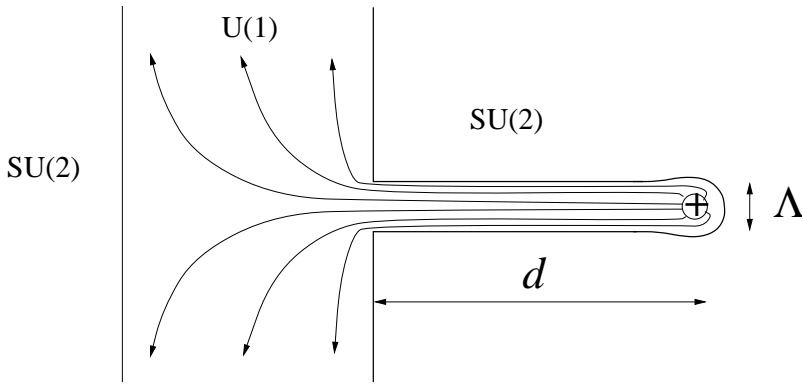}
\smallskip
\parbox{5in}{\caption{\it A test charge which is moved a distance
$d$ off the wall drags a flux tube of thickness $\Lambda$
behind it because electric charge is confined in the bulk.
}}
\end{figure}

We now explore this analogy with D-branes further by considering
what happens when we bring two or more of these walls close
together. For concreteness, let us take a case with $SU(2)$
Higgsed to a $U(1)$ in two regions ${\cal W}_1 = -l_1 < z < 0$
and ${\cal W}_2 = d < z < d + l_2$.
Let us first consider
the case where the walls are very well separated $d >> l_1,l_2$.
Then at distances longer than $l_{1,2}$ we have two $(2+1)$-d walls
with two separate $U(1)'s$ localized on them. To see that there
are really two $U(1)'s$, simply note that the electric field lines
emanating from a charge on ${\cal W}_1$ can never end on a charge
in ${\cal W}_2$ because of the confining region separating them.
Let us further simplify our description by working in the
effective theory at distances $>> d$, where the separation between
the walls cannot be discerned. This is then a $(2+1)-$d theory
with a $U(1) \times U(1)$ gauge group. For the case of $N$
well-separated walls, this very long distance theory has a $U(1)^N$
gauge symmetry.

Next consider the opposite extreme when two walls are
sitting very close to each other $d << 1/\Lambda < l_{1,2}$.
At long distances this case is
indistinguishable from having just one wall with thickness
$(l_1 + l_2)$, and we only localize a single $U(1)$
gauge field in the very long distance theory. Therefore,
as the walls are brought close together, the long-distance
theory sees a ${\it reduction}$ of the gauge symmetry from $U(1) \times U(1)$
to $U(1)$. This is opposite to the D-brane case, where the gauge
symmetry gets enhanced from $U(1) \times U(1) \to U(2)$.
Nevertheless, as we will see below, the physics of the two
situations is very similar.

Let us first try to understand what is going on purely in the
long-distance theory. As the parameter $d$ in the theory is
varied, we go from having a $U(1) \times U(1)$ symmetry for $d >> l_{1,2}$
to just a $U(1)$ symmetry for $d=0$. The most plausible
interpretation is that the $U(1) \times U(1)$ symmetry is Higgsed
somewhere in the transition where $d \sim \Lambda$. Since neither
of the walls is special, we expect that $U(1) \times U(1)$ must be
broken to the diagonal $U(1)$. This satisfies an interesting
consistency check. From the microscopic viewpoint,
when the walls merge to give a new wall of
thickness $(l_1 + l_2)$, the (2+1)-d coupling of the single $U(1)$
should be
\beq
\frac{1}{g^2_3} = \frac{(l_1 + l_2)}{g^2_4}
\eeq
On the other hand, the gauge coupling determined by Higgsing $U(1) \times
U(1)$
to the diagonal subgroup is
\beq
\frac{1}{g^2_{3 diag}} = \frac{1}{g^2_{3,1}} + \frac{1}{g^2_{3,2}}
= \frac{l_1}{g^2_4} + \frac{l_2}{g^2_4}
\eeq
as required.

Therefore, purely from considerations of the very low-energy
theory, we conclude that some new state becomes light when
$d \sim \Lambda$, and acquires a condensate to spontaneously break
$U(1)_1 \times U(1)_2 \to U(1)_{diag}$. The condensate must of course
be a Lorentz scalar, and must be charged under both $U(1)'s$ to
break to the diagonal subgroup. The simplest possibility is that
as $d$ is reduced and becomes smaller than $\sim \Lambda$
a scalar field $\phi^{+,-}$ of charge $(+,-)$ under
$U(1)_1 \times U(1)_2$
becomes light and then tachyonic,
triggering the non-zero condensate $\langle \phi^{+,-}
\rangle$.

We stress that the existence of such a condensate was deduced
by the requirement of a consistent low-energy effective theory.
But we can easily identify a natural candidate for $\phi^{+,-}$ in
the microscopic theory. For $d >> l$, there is a stable
configuration corresponding to the QCD string stretching
between the walls as shown in figure 3.

% FIGURE 3
\begin{figure}[ht]
\centering
\epsfysize=3.in\epsfbox{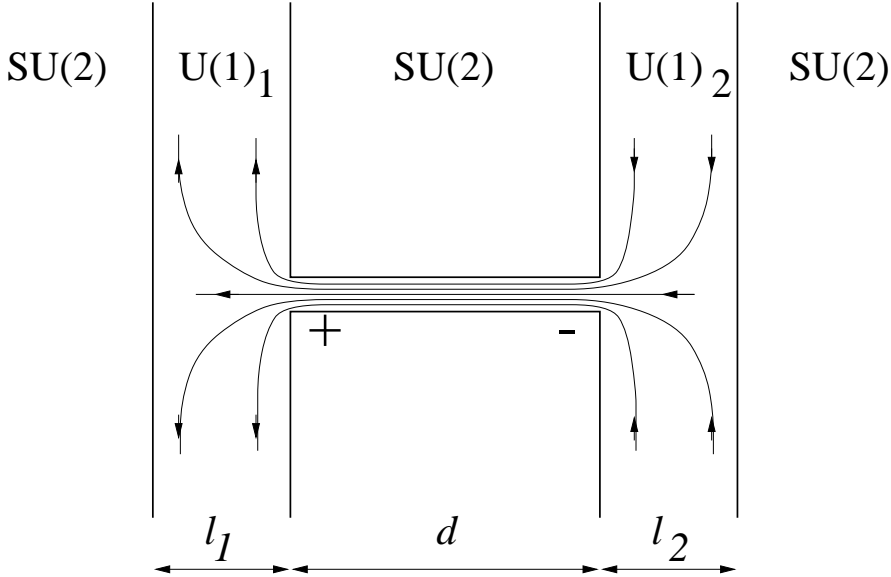}
\smallskip
\parbox{5in}{\caption{\it A color electric flux tube connecting two
walls. In terms of the low energy effective theory on the walls this
flux tube is described by a field charged under both $U(1)$'s.
}}
\end{figure}
One can imagine forming this string as follows.
Place very heavy test quarks $q,\bar{q}$ inside the confining medium
between the walls; a QCD string of confined color electric flux
will stretch between them.
Now, move $q (\bar{q})$ until it is just inside region ${\cal W}_{1(2)}$.
This will cost a great deal of energy $\sim \Lambda^2 d$,
but the resulting string
is stable: it can not break since there are
no dynamical quark states to pop out of the vacuum. Once $q,\bar{q}$
are inside their respective walls, they are no longer in a
confining medium and can be moved off to large distances
\footnote{Of course since the $(2+1)$-dimensional theory is itself
confining at much longer distances, the energy to move $q,\bar{q}$
to infinity diverges logarithmically in the IR.}. The resulting
configuration is just a flux tube with field lines coming
in from infinity on ${\cal W}_1$, through the tube and back out to infinity on
${\cal W}_2$ (see figure 3).
Note that an observer on ${\cal W}_1$ sees this as a state of
charge $+1$ under $U(1)_1$, while his friend on ${\cal W}_2$ measures
this same state to have charge $-1$ under $U(1)_2$.

Thus at least for
$ d >> \Lambda$, we have identified a stable state $\phi^{+,-}$, the
lowest scalar vibrational mode of the QCD string stretching between the walls,
with the quantum numbers we are after. Furthermore, it is clear that
for large $d$ the mass of this mode decreases as $\Lambda^2 d$
as the walls are brought closer. It is now tempting to
speculate that as the walls come very close together, this state gets
lighter and lighter until it becomes tachyonic somewhere around
$d \sim \Lambda$ and condenses. In other words, we imagine that the
mass for $\phi$ as a function of $d$ has the form
\beq
m^2_{\phi}(d) = - c_0 \Lambda^2 + c_1 (\Lambda ^2 d)^2
\eeq
where $c_0,c_1$ are $O(1)$ constants. The contribution
proportional to $c_1$ has a classical origin and dominates when the
string is long, while the first term reflects a (presumably
quantum-mechanical) tachyonic
instability of the unstretched QCD string. Of course, since the
flux tube has a thickness $O(\Lambda)$, in the interesting region
it is as long as it is thick so a ``string" picture is
necessarily heuristic. In terms of the microscopic $(3+1)$-d
description the condensate of the tachyonic scalar can be understood
as a spontaneous deconfinement transition of the QCD vacuum
between the two walls due to a condensate of flux tubes. 

It is amusing that while the end result of bringing these
``branes" together is very different from the case of bringing D-branes
together, the physics has a similar interpretation: the strings
stretching between the branes become light and donate their lowest
excitations to the effective theory. In the case of D-branes,
the lowest-lying excitations of open strings contain gauge fields
which enhance the gauge symmetry. On the other hand, our
flux-strings have no massless gauge fields so
there is no enhancement of gauge symmetry.
Instead, the lightest excitation that is donated is tachyonic
and further breaks the gauge group! It is also interesting that
the tachyonic instability of the QCD string here does not imply
that the theory is sick and should be discarded; it simply means
that the correct vacuum, where the strings have condensed, must be
chosen.

Before we move on note that we can obtain some
information about the dynamics of flux tubes
by simply translating $(2+1)$ dimensional results into our
microscopic description. For example, the fact that the
$(2+1)$-d $U(1) \times U(1)$ theory confines tells us that flux
tubes in our picture are confined. A stable finite energy
configuration is a spinning bound state of two flux tubes
of opposite flux. In the case of large wall separation when
the flux tubes are long and heavy, this is a non-relativistic
bound state but as we tune the distance between the two walls
such that the scalar mode $\phi^{+,-}$ becomes light the
bound state becomes non-relativistic. It is amusing that these
``bound states'' of flux are spinning closed flux-strings
which overlap both walls.
 
We will not discuss at length an obvious generalization to $N$ walls with
flux tubes stretching between any pair of neighboring walls. These
strings are charged under neighboring $U(1)$ and donate the
necessary scalar fields to break $U(1)^N \rightarrow U(1)_{diag}$
as all $N$ walls are merged. 

Instead, we cannot resist the temptation to describe an
interesting generalization with strings corresponding to baryons
of the confined gauge theory in the bulk. 
First note that the above construction of domain walls with
trapped gauge fields generalizes to branes of dimension $(1+1)$.
To construct such a ``string'' or ``1-brane''
consider a patch $\cal W$ in the $y-z$
plane where the bulk non-abelian gauge group is broken to
$U(1)$ as depicted in figure 4.

% FIGURE 4
\begin{figure}[ht]
\centering
\epsfysize=2.in\epsfbox{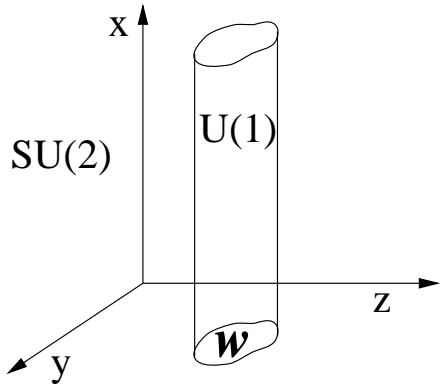}
\smallskip
\parbox{5in}{\caption{\it A photon can be localized on a 1+1
dimensional ``1-brane'' by embedding the $U(1)$ into a confining
$SU(2)$ in the bulk.
}}
\end{figure}

Again, the bulk non-abelian gauge theory
is chosen to confine at distances $\Lambda^{-1}$
taken to be much shorter than
the square root of the area $A$ of $\cal W$. This traps
a $(1+1)$ dimensional $U(1)$ gauge theory with gauge coupling
$g_2^2\sim g_4^2/A$ on the string. Given two such 1-branes with
areas $A_{1,2}$ and associated $U(1)$ gauge theories one can
consider bringing the two 1-branes in contact. Again, the
low energy theory sees a reduction of the gauge group from $U(1)\times U(1)$
to $U(1)_{diag}$ with the interpretation of Higgsing of the gauge
group via a scalar which becomes light and tachyonic as the two regions
are brought within distances of order $\Lambda^{-1}$. Evidence for
this interpretation is the matching of $U(1)$ couplings which in
this case reads
\beq
\frac{1}{g^2_{2 diag}} = \frac{(A_1+A_2)}{g_4^2} =
\frac{A_1}{g^2_4} + \frac{A_2}{g^2_4} =
\frac{1}{g^2_{2,1}} + \frac{1}{g^2_{2,2}}\ .
\eeq
Just as
in the case of domain walls a QCD string connecting $A_1$ with $A_2$
has the correct quantum numbers to supply this scalar.

Consider now a situation with a confining gauge group $SU(N)$
in the bulk and $N$ patches with roughly equal areas
$A_i$ for ${i=1,..,N}$ on which
the running of the $SU(N)$ coupling has been slowed down.
(For example, this could be arranged by adding matter fields in the
adjoint representation of $SU(N)$ which have very large masses in
the bulk but are light on the 1-branes.)
Then the four dimensional gauge
coupling remains small on the $N$ patches, and below the matching
scale $\mu \sim A_i^{-1/2}$ the long distance physics is described
by a $(1+1)$ dimensional $SU(N)^N$ gauge theory.
As before we can create flux tubes connecting any pair of
the various 1-branes by placing a pair of heavy test quarks
$q$ and $\bar{q}$ in the confined bulk, pulling them apart to form
a flux string and then moving the two quarks in two separate regions
$A_i$ and $A_j$. After removal of the test quarks we are left with
a flux tube which transforms in the fundamental representation
of $SU(N)_i$ and an antifundamental of $SU(N)_j$.

What happens if we start with $N$ test quarks in a color singlet
state corresponding to a baryon of the $SU(N)$ bulk gauge group?
We can now move each of the test quarks into a different one
of the patches $A_i$; the flux tubes from each of these quarks
meet at a common junction in the bulk where $N$ units of
flux combine into a color singlet. After removal of the test
quarks we are left with a baryonic $N$-pronged flux tube
connecting the $N$ regions as in figure 5.

% FIGURE 5
\begin{figure}[ht]
\centering
\epsfysize=2.in\epsfbox{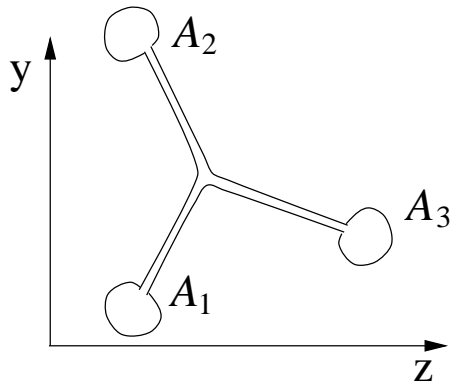}
\smallskip
\parbox{5in}{\caption{\it A 3-pronged baryonic flux tube connecting
three 1-branes.
}}
\end{figure}

This baryonic string is stable and
has a mass of order $N \Lambda^2$ times the characteristic
distance between the various patches. In the low energy
effective $SU(N)^N$ theory its lowest vibrational mode
would be described by a massive field which transforms in the fundamental
representation of each of the $SU(N)$'s. As we bring all of the
$N$ patches close together, the baryonic string as well as
all the ``mesonic'' strings become light. 
As we bring the 1-branes very close to each other we expect
a condensate of strings which deconfines the vacuum in the
region between the 1-branes, causing them to merge. In
the long distance theory this is described by
a condensate of scalars which breaks $SU(N)^N \rightarrow SU(N)$.

\section*{Acknowledgements} We thank Michael Peskin and Eva
Silverstein for discussions.
Our work is supported by the Department of Energy under contract
DE-AC03-76SF00515.

\section*{Appendix}
In this appendix we formalize the conclusions regarding the trapping
massless electric or magnetic photons 
in the theory where a $U(1)$ gauge field is higgsed
away from ${\cal W}$ but is unbroken inside ${\cal W}$.
Let us consider formally the limit
where $M l \to \infty$, so that the photon outside is
really infinitely heavy. The $U(1)$ then really only exists
inside ${\cal W}$; with a Lagrangian given by
\beq
{\cal L} = \int d^3 x \int_{-l/2}^{l/2} dz
\frac{1}{g^2_4} F^{\mu \nu} F_{\mu \nu}.
\eeq
This just looks like the compactification of a $U(1)$ gauge
theory from $4 \to 3$ dimensions on an interval of length $l$.
However, the spectrum of the theory at energies beneath
the compactification scale $l^{-1}$ depends crucially on the
boundary conditions
imposed on $F^{\mu \nu}$ at $z=+l/2,-l/2$. This is because
massless states in the low energy theory must be zero modes
in the $z$ direction, 
and are therefore sensitive to the boundary conditions, which
may or may not eliminate them.
In the present case, the region outside the wall is superconducting, so the 
appropriate boundary conditions are that the electric field is perpendicular to
the wall (true for any conductor) and that the magnetic field is parallel to 
the wall (which is only true for a superconductor); that is       
\beq
E_x=E_y=0,\quad B_z=0
\eeq
which can be written more covariantly as 
\beq
F^{ab} = 0, \quad a,b=t,x,y.
\eeq
This makes it clear that a massless photon coupling to electric charge is not
present in the low energy theory, it is projected out of the usual Kaluza-Klein
spectrum by the boundary conditions. The usual KK scalar, corresponding to 
$F^{a3}$, remains in the massless spectrum, but does not couple to electric charge. Rather, a massless magnetic photon has been trapped. Indeed, the boundary
conditions can also be written as
\beq
\tilde{F}^{a3} = 0
\eeq
which leaves the zero mode of $\tilde{F}^{ab}$ in the massless spectrum. 
This of course works because in $(2+1)$ dimensions, a scalar is dual to a vector
field. 

Of course in both cases, we also have a tower of massive states.
This follows from the  standard Kaluza-Klein analysis with the boundary conditions
appropriately imposed. But we can also see it in another way. Let us generalize to the case 
of an $n$ dimensional wall of thickness $l$ in an $n+1$ dimensional space. Let $\vec x$ be the $n$-dimensional
coordinates, and $y$ the $n+1$'th coordinate. Placing an electric charge at the origin, let us compute the
electric potential at the point $\vec x,y=0$ on the wall. We can enforce the boundary conditions 
by placing an infinite sequence of image charges of charge $(-1)^q$ at $\vec x=0,y=q l$. The potential is then
\beq
V(\vec x) = \sum_{q = -\infty}^{+\infty}  \int \frac{d^n k}{(2 \pi)^n} \frac{d k'}{(2 \pi)} \frac{e^{i (\vec k \vec x +  k' l q - \pi q)}}{\vec k^2 + k'^2}
\label{pot}
\eeq   
where we have used the expression for the $(n+1)$ dimensional Coulomb potential in terms of its Fourier transform.
If we now use the familiar Poisson resummation identity
\beq
\sum_{q=-\infty}^{\infty} e^{i q \theta} = 2 \pi \sum_{s = -\infty}^{\infty} \delta(\theta - 2 \pi s)
\eeq
we can perform the integral over $k'$, leaving 
\beq
V(\vec x) = \sum_{s = -\infty}^{\infty} \int \frac{d^n k}{(2 \pi)^n} \frac{e^{i \vec k \vec x}}{\vec k^2 + (2 \pi/l)^2(s + 1/2)^2}.
\eeq
Note that the integrand is just the $n$-dimensional Yukawa potential for a field of mass $2 \pi/l \times (s+1/2)$. Therefore,
we have shown that the potential can be expressed in terms of a sum over a tower of $n$ dimensional massive states. 
If we instead place a magnetic charge at the origin, an infinite sequence of
magnetic image charges of the {\it same} sign enforce the boundary condition,
and the $\pi q$ term in eqn.(\ref{pot}) disappears. Therefore, the potential is due to a sum over $n$ dimensional massive fields of mass
$2 \pi /l \times s$, which for $s=0$ includes the massless magnetic photon we expect in this case.

\newcommand{\cm}{Commun.\ Math.\ Phys.~}
\newcommand{\prl}{Phys.\ Rev.\ Lett.~}
\newcommand{\pr}{Phys.\ Rev.\ D~}
\newcommand{\pl}{Phys.\ Lett.\ B~}
\newcommand{\ibar}{\bar{\imath}}
\newcommand{\jbar}{\bar{\jmath}}
\newcommand{\np}{Nucl.\ Phys.\ B~}
\def\pl#1#2#3{{\it Phys. Lett. }{\bf B#1~}(19#2)~#3}
\def\zp#1#2#3{{\it Z. Phys. }{\bf C#1~}(19#2)~#3}
\def\prl#1#2#3{{\it Phys. Rev. Lett. }{\bf #1~}(19#2)~#3}
\def\rmp#1#2#3{{\it Rev. Mod. Phys. }{\bf #1~}(19#2)~#3}
\def\prep#1#2#3{{\it Phys. Rep. }{\bf #1~}(19#2)~#3}
\def\pr#1#2#3{{\it Phys. Rev. }{\bf D#1~}(19#2)~#3}
\def\np#1#2#3{{\it Nucl. Phys. }{\bf B#1~}(19#2)~#3}
\def\mpl#1#2#3{{\it Mod. Phys. Lett. }{\bf #1~}(19#2)~#3}
\def\arnps#1#2#3{{\it Annu. Rev. Nucl. Part. Sci. }{\bf #1~}(19#2)~#3}
\def\sjnp#1#2#3{{\it Sov. J. Nucl. Phys. }{\bf #1~}(19#2)~#3}
\def\jetp#1#2#3{{\it JETP Lett. }{\bf #1~}(19#2)~#3}
\def\app#1#2#3{{\it Acta Phys. Polon. }{\bf #1~}(19#2)~#3}
\def\rnc#1#2#3{{\it Riv. Nuovo Cim. }{\bf #1~}(19#2)~#3}
\def\ap#1#2#3{{\it Ann. Phys. }{\bf #1~}(19#2)~#3}
\def\ptp#1#2#3{{\it Prog. Theor. Phys. }{\bf #1~}(19#2)~#3}

\end{document}